\documentstyle[12pt]{article}
\begin{document}
\input epsf

\begin{center}
\large{{\bf Consequences on some dark energy-candidates from the type
Ia supernova SN 1997ff}}
\end{center}

\begin{center}
R. G. Vishwakarma\footnote{E-mail: vishwa@iucaa.ernet.in}\\


{\it Inter-University Centre for Astronomy \& Astrophysics,
 Post Bag 4, Ganeshkhind, Pune 411 007, India }
\end{center}


\noindent
{\bf ABSTRACT:}\\
We examine the status of various {\it dark energy}-models in light of the
recently observed SN 1997ff
at $z \approx 1.7$. The modified data still fit a pure cosmological constant $\Lambda$
or a quintessence with an equation of state similar to that of $\Lambda$. The
kinematical $\Lambda$-models, $\Lambda\sim S^{-2}$ and $\Lambda\sim H^2$, also fit
the data reasonably well and require less {\it dark energy} density (hence more
matter energy density) than is required by the constant $\Lambda$-model. However,
the model $\Lambda\sim S^{-2}$ with low energy density becomes unphysical as it
cannot accommodate higher redshift objects.

We also examine an alternative explanation of the data, viz., the absorption
by the intervening whisker-like dust and find that the quasi-steady state
model and the FRW model $\Omega_{\rm m0}=0.33$, {\it without any dark
energy}
also fit the data reasonably well.

We notice that the addition of SN 1997ff to the old data has worsened
the fit to most of the models, except a closed FRW model with a constant $\Lambda$
and a closed {\it quintessence}-model with $\omega_\phi=-0.82$, and the models
have started departing from each other as we go above $z=1$. However,
to make a clear discrimination possible, a few more supernovae with $z>1$ are
required.

We have also calculated the age of the universe in these models and find
that, in the models with a constant $\Lambda$, the expansion age is
uncomfortably close to the age of the globular clusters. {\it Quintessence}-models show
even lower age. The kinematical $\Lambda$-models are, however, interesting in this
connection (especially the model $\Lambda\sim H^2$), which give remarkably large
age of the universe.

\vspace{.2cm}
\noindent
{\it Subject heading:} cosmology: theory, SN observations, dark energy.

\newpage
\noindent
{\bf 1. INTRODUCTION}

\vspace{.2cm}
\noindent
The recent luminosity-redshift observations of type Ia supernovae (Riess
et al, 1998; Perlmutter et al, 1999) reveal that the dominant constituent of the
universe is some weird form of matter with negative
pressure, commonly referred to as `{\it dark energy}', which gives rise to an
accelerated expansion and thus explains the unexpected faintness of the
supernovae. This mysterious cosmic matter is different from the cold, collisionless,
nonbaryonic dark matter (CDM) (evidenced by the dynamical measurements of the
total mass density) which dominates the gravitationally clustered matter and
clumps like the ordinary matter. The  {\it dark energy}, on the contrary, does not
clump but repels matter instead, due to its negative pressure. It is `{\it dark}'
because it is not recognizable by direct observations and interacts with matter
via gravity only. Existence of {\it dark energy} is also suggested by the
anisotropy measurements of the microwave background radiation (MBR) which indicate that
the universe is flat (Dodelson \& Knox, 2000; Jaffe et al, 2001; Pryke et al,
2001; Netterfield et al, 2001).
As we also have evidence that there is not enough matter,
including dark matter, to account for the critical density
(Turner, 2001), this provides
evidence for an additional component of energy $-$ the {\it dark energy}.
It has been shown that the presence of {\it dark energy} in a CDM-model
yields results that fit better with the observations of large scale structure
(Kofman et al, 1993; Krauss \& Turner, 1995; Ostriker \& Steinhardt, 1995;
Bertschinger, 1998; Bahcall et al, 1999; Sahni \& Starobinsky, 2000).

   An obvious candidate for {\it dark energy} is a positive cosmological constant
$\Lambda$ which represents the energy density of the quantum vacuum
($\rho_{\rm v}=\Lambda/8\pi G$). However,
a pure cosmological constant cannot explain the small vacuum energy density of
$10^{-120}$ M$_{\rm Pl}^4$ without running into fine tuning problems.
Besides this old cosmological constant problem to
understand why the vacuum energy density is so small, now
a new problem, known as the {\it coincidence problem}, is to understand why
the vacuum energy density is comparable to the present mass density of the
universe. That is, why we happen to live in this special epoch.

  {\it Quintessence} and kinematical $\Lambda$, which are other popular options for
{\it dark energy}, have been proposed to solve these problems and have been
extensively studied in the literature (for a recent review, see Sahni \&
Starobinsky, 2000). Brane-world scenarios, resulting from the hidden additional
space dimensions, have recently been proposed as an alternative to the
{\it dark energy} in explaining the current acceleration of the universe.
These theories are motivated by the fact that they are realized in string theory
which is a promising candidate for quantum gravity.
However, by considering one of such models, it has been shown recently that
the model is disfavoured by the existing cosmological datasets (Avelino \& Martins, 2001).
We hope that other brane-world models will be brought into the realm of
cosmological testability in the near future.

   Supernovae Ia, which are thought to be thermonuclear explosions of
carbon oxygen white dwarfs (Trimble, 1983; Woosley \& Weaver, 1986), are almost
universally regarded as {\it standard candles} due to the following reasons.
The variability of the luminosity of type Ia supernovae is small and an empirical
relation exists between the decline rate of their observed luminosity and the
absolute luminosity (de Vaucouleurs \& Pence, 1976; Pskovskii 1977; Branch 1981,
1982; Boisseau \& Wheeler 1991).
That is, a brighter supernova fades away more slowly than
the one which is less bright. Moreover, they have high enough absolute luminosity
to make it possible to observe them at cosmological scales.

As the {\it luminosity distance} $d_{\rm L}$ depends
sensitively on the spatial curvature and the expansion dynamics
of the models, the magnitude-redshift ($m$-$z$) relation for the
high redshift supernovae has been proposed as a potential
test for cosmological models and thus provides a useful tool
to enable a discrimination to be made between models of various
{\it dark energy}-candidates.
Very recently, a supernova of type Ia (SN 1997ff) at $z \approx 1.7$ (the highest redshift
SN observed so far) has been identified by the {\it Hubble Space
Telescope} (Riess et al, 2001).
It would be worthwhile to study the consequences of this event on various
{\it dark energy}-candidates.

In this paper, we study the consequences on various
{\it dark energy}-candidates from the modified data, by considering the constant
$\Lambda$ FRW model and some variable $\Lambda$-models of dynamical
and kinematical origin.
The claim, that the observed dimming of the supernovae can, alternatively, be
explained by the absorption from the whisker-like dust (Aguire, 1999; Banerjee
et al, 2000), is also
examined in the standard Friedmann model (with a vanishing $\Lambda$) and the
quasi-steady state model.

   We would like to mention that there are a number of possible contaminants
to the measurements of SN 1997ff, for example, the host extinction and
the magnification of the supernova by gravitational lensing due to the presence
of galaxies lying along our line of sight (Moertsell, Gunnarsson \& Goobar, 2001;
Lewis \& Ibata, 2001). If these effects are manifested, they could significantly
reduce the cosmological utility of this supernova, and of high redshift supernovae
in general. However, in this analysis, we consider the measurements at their
`face values', which is also motivated by the tangential shear test performed
and discussed by Riess et al (2001).

   Throughout our analysis, we shall be considering the Robertson-Walker (R-W)
metric
\begin{equation}
ds^2=-dt^2 + S^2(t) \left\{\frac{dr^2}{1-kr^2} +
r^2(d\theta^2+\sin^2\theta d\phi^2)\right\},
\end{equation}
in which case, the Einstein field equations lead to

\begin{equation}
\frac{\dot S^2}{S^2}+\frac{k}{S^2}=\frac{8\pi
G}{3}\rho _{\mbox{{\scriptsize tot}}},
\end{equation}

\begin{equation}
-\frac{\ddot S}{S}=\frac{4\pi
G}{3}\left(\rho _{\mbox{{\scriptsize tot}}}+3p _{\mbox{{\scriptsize tot}}}\right),
\end{equation}
where $\rho_{\rm tot}\equiv\rho_{\rm m}+\rho_{\rm dark ~energy}$ and
$p_{\rm tot}\equiv p_{\rm m}+p_{\rm dark ~energy}$ with
$\rho_{\rm m}\equiv\rho_{\rm rest ~mass}+\rho_{\rm radiation}+
\rho_{\rm dark ~matter}$ and  $p_{\rm m}= p_{\rm radiation}$.

\vspace{1cm}
\noindent
{\bf 2. QUINTESSENTIAL MODEL}

\vspace{.2cm}
\noindent
It is increasingly popular to consider the possibility that the vacuum energy
is not constant but rather evolves with the expansion of the universe. For
example, a slowly varying scalar field $\phi$ (commonly known as
{\it quintessence}) with a suitable potential $V(\phi)$ coupled to matter through
gravity, contributes a vacuum energy density $\rho_{\phi}$ and a pressure
$p_\phi$ as

\begin{equation}
\rho_{\phi}=\frac{1}{2}\dot\phi^2 + V(\phi), ~ ~
p_{\rm \phi}=\frac{1}{2}\dot\phi^2 - V(\phi),
\end{equation}
with an equation of state $\omega_\phi\equiv p_\phi/\rho_\phi$ which may be
a function of time ($\omega_\phi=-1$ corresponds to a pure cosmological
constant). One notes that if the field $\phi$ is evolving sufficiently slowly
so that $\dot\phi^2< V(\phi)$, one can have $\omega_\phi<0$. The condition
$(\rho +3p)_{\rm tot}<0$ for an accelerating expansion (see equation (3)) can
then be satisfied.

As the scalar field interacts minimally with matter and radiation, these are
conserved separately. This leads to

\begin{equation}
\rho_{\rm m}\sim S^{-3(1+\omega)}\sim (1+z)^{3(1+\omega)}, ~ ~ ~\rho_\phi \sim
S^{-3(1+\omega_\phi)}\sim (1+z)^{3(1+\omega_\phi)},
\end{equation}
where $\omega=0$ and $1/3$ for, respectively, rest mass (dust) and radiation,
as usual. In deriving (5), we have assumed that $\omega_\phi$ is a constant
which is a reasonable
approximation as far as the equation of state changes slowly with time.
We have also assumed that the radiation energy is negligible for
$(1+z)<<10^3$. It is clear that $\rho_\phi$ decays slower than $\rho_{\rm m}$
as the potential energy  $V(\phi)$ starts dominating over the kinetic energy
$\dot\phi^2/2$ of the scalar field and $\omega_\phi$ turns negative. However,
in the earlier epoch (with $\rho_{\rm rad} \sim S^{-4}$) where the kinetic
energy is dominant, $\rho_\phi$ decays faster (as fast as $S^{-6}$) than
$\rho_{\rm rad}$. Equation (5) implies that, in the present phase of evolution,
\begin{equation}
\rho_\phi \frac{<}{>} \rho_{\rm m} ~ ~ \mbox{according as} ~ ~
(1+z) \frac{>}{<} \left(\frac{\Omega_{\phi 0}}{\Omega_{\rm m 0}}\right)^{
-1/3\omega_\phi}, ~ ~ \Omega_{\phi}>0,
\end{equation}
where $\Omega_i$'s are, as usual, the energy densities of the respective components
in units of the critical density (i.e., $\Omega_i\equiv 8\pi G \rho_i/3 H^2$)
and the subscript `0' denotes the value of the quantity at the present epoch.
Thus in a flat model with $\Omega_{\rm m 0}=0.34$ (inferred from the SN data, as
we shall see later), the redshift ($\equiv z_{\rm eq}$) at which $\rho_\phi$
started dominating over $\rho_{\rm m}$, comes out as $z_{\rm eq}= 0.25$ if
the {\it dark energy} is represented by a pure cosmological constant.

From equations (2), (3) and (5), it is not difficult to calculate the evolution
of the deceleration parameter $q\equiv -\ddot S/SH^2$ in the present phase of
evolution:

\begin{equation}
q(z)=\frac{1}{2}\left[\frac{1+(\Omega_{\phi 0}/\Omega_{\rm m 0})(1+3\omega_\phi)
(1+z)^{3\omega_\phi}}
{1+(\Omega_{\phi 0}/\Omega_{\rm m 0})(1+z)^{3\omega_\phi}-
(\Omega_{k0}/\Omega_{\rm m 0})(1+z)^{-1}} \right],
\end{equation}
where $\Omega_k\equiv k/S^2 H^2$. From this, it follows that for $\omega_\phi<-1/3$
 and $\Omega_\phi>0$,

\begin{equation}
q \frac{<}{>}0 ~ ~ \mbox{according as} ~ ~ (1+z) \frac{<}{>} \left[
(3\mid\omega_\phi\mid-1)\frac{\Omega_{\phi 0}}{\Omega_{\rm m 0}}\right]^{
1/3\mid\omega_\phi\mid},
\end{equation}
in flat or open models. This implies that the universe must have been
decelerating in the past if it is accelerating at present. For example, in a flat
universe with the {\it dark energy} in form of a constant $\Lambda$,
the value $\Omega_{\rm m 0}=0.34$ (which guarantees an acceleration at the present
epoch), implies that the expansion of the
universe shifted from deceleration to acceleration at $z=0.57$. This happened
earlier than the epoch when the vacuum energy density took over to the
matter energy density.

In order to fit the models to the supernovae data, we now derive the magnitude
($m$)$-$redshift ($z$) relation. The {\it luminosity distance} $d_{\rm L}$ of a
source with redshift $z$ is given by

\begin{equation}
d_{\rm L} (z)=\frac{(1+z)}{H_0 \sqrt{{\cal K}}}\xi\left(\sqrt{{\cal K}}
H_0 \int_0^z \frac{{\rm d} z'}{H(z')}\right),
\end{equation}
where

$\xi(x)=\sin (x)$ ~ ~ with ~ ~ ${\cal K}=\Omega_{k0}$ ~ ~ ~when ~ ~
$\Omega_{k0}>0$,

$\xi(x)=\sinh (x)$  ~ ~with ~ ~${\cal K}=-\Omega_{k0}$~ ~ when ~ ~
$\Omega_{k0}<0$,

$\xi(x)=x$ ~ ~ ~ ~ ~ with ~ ~ ~${\cal K}=1$ ~ ~ ~ ~when ~ ~
$\Omega_{k0}=0$.

\noindent
The apparent magnitude $m$ of the source is, then, given by

\begin{equation}
m(z)={\cal M} + 5 \log_{10} (H_0 d_{\rm L}),
\end{equation}
where ${\cal M}\equiv M-5\log_{10}H_0+25$, $M$ is the absolute luminosity of the
source and $d_{\rm L}$ is measured in Mpc. It is worth noting that in deriving
equations (9) and (10), the only assumption one needs to make, is that of the
validity of the R-W metric given by equation (1).
Hence these equations are applicable to all the homogeneous and isotropic models.
The additional knowledge of the evolution of the Hubble parameter $H(z)$ in a
particular model,
determines the $m$-$z$ relation completely in the model. The expression
for $H(z)$ in this model follows from equations (2) and (5) as
\begin{equation}
H(z)=H_0[\Omega_{\rm m 0} (1+z)^3 +\Omega_{\phi 0}(1+z)^{3(1+\omega_\phi)}
-\Omega_{k0}(1+z)^2]^{1/2}.
\end{equation}

We consider the data on the redshift and effective magnitude of SN 1997ff
together with the sample of 54 supernovae from the Perlmutter et al' primary
fit C. There are two ways Riess et al (2001) have measured SN 1997ff.
One is to use the photometric redshift (1.7) of the supernova, in which case
the resulting distance and redshift is a complex confidence contour (see figures
10 and 11 in their paper). The other comes from the deployment of the spectroscopic
redshift (1.755) of the host which has the virtue of being much more accurate,
but is subject to being in error if the line one identifies is not correct.
However, both methods give a very similar result and a good representation of the
data is $z=1.755\pm0.05$, $m^{\rm eff}=25.68\pm 0.34$ (from a personal
discussion with Adam G. Riess).
The best-fitting parameters are estimated by minimizing the function
\begin{equation}
\chi^2=\sum_{i=1}^{55} \left[\frac{m_i^{\rm eff} - m(z_i)}
{\sigma_{m_i^{\rm eff}}}\right]^2,
\end{equation}
where $m_i^{\rm eff}$ refers to the effective
magnitude of the $i$th supernovae which has been corrected by the lightcurve
width-luminosity correction, galactic extinction and the K-correction from
the differences of the R- and B-band filters and the dispersion
$\sigma_{m_i^{\rm eff}}$ is the uncertainty in $m_i^{\rm eff}$.

Before fitting the data to the model, we recall that any source
of energy and momentum should satisfy the {\it null energy condition} which,
in the R-W metric, is equivalent to requiring $\rho+p\geq 0$. In the case of
{\it quintessence}, this condition leads to $\rho_\phi(1+\omega_\phi)\geq 0$.
The global best-fitting model, where $\chi^2$ is minimized by giving free rein
to $\Omega_{\rm m 0}$, $\Omega_{\Lambda 0}$, $\omega_\phi$ and ${\cal M}$, i.e.
at 51 degrees of freedom (dof), is obtained
as $\Omega_{\rm m 0}=0.71$, $\Omega_{\phi 0}=1.79$ and $\omega_\phi=-0.82$
with $\chi^2=56.82$, which is indeed a good fit ($\Delta\chi^2\equiv\chi^2/$dof
$=1.12$). The global
best fit for a constant $\Lambda$ ($\omega_\phi=-1$) also gives a similar value,
viz., $\chi^2=56.89$ (at 52 dof) for the model
$\Omega_{\rm m 0}=0.87$, $\Omega_{\Lambda 0}=1.51$, which also shows a very good
fit ($\Delta\chi^2=1.09$).
We note that without including SN 1997ff, the global best-fitting quintessence
model was obtained as $\Omega_{\rm m 0}=0.66$, $\Omega_{\phi 0}=2.02$ and
$\omega_\phi=-0.75$ with $\chi^2=56.81$ (at 50 dof, i.e., $\Delta\chi^2=1.14$).
The global best-fitting constant $\Lambda$-model was: $\Omega_{\rm m 0}=0.79$,
$\Omega_{\Lambda 0}=1.4$ with $\chi^2=56.85$ (at 51 dof, i.e.,
$\Delta\chi^2=1.11$).

We notice that the estimates of $\Omega_{\rm m 0}$,
mentioned above, are rather higher than its value estimated from various
measurements, most of which seem to indicate a low value
$\Omega_{\rm m 0}\approx 0.33$ (Turner, 2001), though, one can still argue that
there is some more matter in
an yet undetected form. Moreover, there are also some measurements which favour
a higher value of $\Omega_{\rm m 0}$ (Schindler, 2001).

If we stick to the flat models ($\Omega_{\rm m}+\Omega_\phi=1$), we find
that the minimum value of
$\chi^2$ decreases as $\omega_\phi$ decreases. Under the constraint of the
{\it null energy condition}, the best-fitting model is obviously that for
$\omega_\phi=-1$ (i.e., for a constant $\Lambda$), which is given by
$\Omega_{\rm m 0}=0.34$ with  $\chi^2=62.01$ (at 53 dof).
This although represents a good fit ($\Delta\chi^2=1.17$), but not as good as
the one with the non-flat models
mentioned above. Nonetheless, the estimated value $\Omega_{\rm m 0}=0.34$ is
remarkably close to the most favoured value $\Omega_{\rm m 0}\approx 0.33$ from
various observations, as mentioned above. The best-fitting flat model, without
SN 1997ff, is $\Omega_{\rm m 0}=0.28$  with  $\chi^2=57.70$ (at 52 dof).
Thus, although the addition of SN 1997ff has worsened the
fit to the flat model considerably, however, the
modified data still fit either a constant $\Lambda$ or a {\it quintessence}
whose equation of state is similar to that of $\Lambda$.
 In the following, we shall limit our analysis to the flat models only,
which are consistent with the MBR observations (which indicate that
$\Omega_{\rm tot}$ is very close to 1).

\vspace{1cm}
\noindent
{\bf 3. PHENOMENOLOGICAL MODELS OF KINEMATICAL $\Lambda$}

\vspace{.2cm}
\noindent
Models of this category, providing another popular candidate for {\it dark energy},
have been proposed to explain the small present value of vacuum energy density through a
kinematical decay of $\Lambda$. The laws for the decay of $\Lambda$ are
phenomenological in nature and follow either from some dimensional arguments
(Chen \& Wu, 1990; Carvalho et al, 1992; Vishwakarma, 2000), or some
symmetry considerations (Ozer \& Taha, 1987; Vishwakarma, 2001a), or simply by
assuming $\Lambda$ as a function of either the cosmic time or any other parameter
of the FRW model (for a brief summary, see, Overduin \& Cooperstock, 1998).
Here we consider the following three most popular models from this category:

$\bullet$ case (1): $\Lambda =n_1 S^{-2}$,

$\bullet$ case (2): $\Lambda =n_2 H^2$,

$\bullet$ case (3): $\Lambda =n_3 t^{-2}$,

\noindent
where the constants $n_i$'s are new cosmological parameters, replacing the original $\Lambda$,
and are to be determined from the observations. The first model is suggested
by a dimensional analysis (Chen \& Wu, 1990) and also results from a contracted
Ricci-collineation
along the fluid flow vector (Vishwakarma, 2001a) and has been the subject of
most attention. The second model has been proposed through the similar dimensional
arguments (Carvalho et al, 1992) and also follows from other arguments.
In view of the present estimates of $\Lambda$ being of the order of $H^2_0$,
this ansatz seems as a natural dynamic law for the decay of $\Lambda$.
The third model has also been considered by several authors by
imposing supplementary conditions which are equivalent to assuming
a power law for the scale factor (Lau, 1985; Bertolami, 1986; Berman \& Som, 1990;
Lopez \& Nanopoulos 1996).  However, with the flat R-W metric, the last two
models (and many other models too, for example, $\Lambda \sim \rho$, see,
Vishwakarma, 2001b) become identical.

\vspace{1cm}
\noindent
{\bf Case (1): $\Lambda =n_1 S^{-2}$}

\vspace{.2cm}
\noindent
The dynamics of the model (for $k=0$) can be obtained from equations (2) and
(3) as

\begin{equation}
\rho_{\rm m}=C_1 S^{-3(1+w)} + \frac{n_1S^{-2}}{4\pi G(1+3w)},
 ~ ~ C_1=\mbox{constant},
\end{equation}

\begin{equation}
\dot{S}^2=\frac{8\pi G C_1}{3} ~ S^{-(1+3w)} + \frac{n_1(1+w)}{(1+3w)}.
\end{equation}
The constants $n_1$ and $C_1$ can be written in the following form

\begin{equation}
n_1=3H_0^2 S_0^2 (1-\Omega_{\rm m 0}), ~ ~ ~ C_1=\frac{3H_0^2 S_0^3}{8\pi G}
(3\Omega_{\rm m0} - 2).
\end{equation}
The value of the constant $C_1$ in the early radiation-dominated epoch can be
calculated, in the same way, in terms of the parameters at, say, the Planck
time or the time of the last scattering. The expression for the Hubble parameter,
in the present phase, yields from equation (14) as

\begin{equation}
H(z)=H_0[(3\Omega_{\rm m 0} -2)(1+z)^3 - 3(1-\Omega_{\rm m 0})(1+z)^2]^{1/2}.
\end{equation}
From this and equation (13), we notice that the accelerating models are severely
constrained. As $q(z)=4\pi GC_1/3 S^3H^2$, it is clear
that for $\Omega_{\rm m 0}<2/3$ (or $\Omega_{\rm m 0}< 2\Omega_{\Lambda 0}$,
in the non-flat models), the models have an accelerating expansion throughout this
phase. However, the maximum redshift of an object
in these models with a positive $\Lambda$, is given by
$z_{\rm m \rm a \rm x}=\Omega_{\rm m 0}/(2-3\Omega_{\rm m 0})$ (in the non-flat
models, $z_{{\rm max}}$ is either $1/(2\Omega_{\Lambda 0}-\Omega_{\rm m 0})$ or
$\Omega_{\rm m 0}/(2\Omega_{\Lambda 0}-\Omega_{\rm m 0})$, whichever is smaller).
Thus the low density models are ruled out by the existence of high redshift
quasars. Nevertheless, the model can increase $z_{{\rm max}}$ considerably by
selecting $\Omega_{\rm m 0}$ sufficiently close to 2/3 (or to
$2\Omega_{\Lambda 0}$ in the non-flat models), if permitted by the observations.
For example, with  $\Omega_{\rm m 0}=0.65$, $z_{{\rm max}}$ becomes as high as 13.
However, the model, then, would have to resort to some other explanation for the
MBR, radically different from
the standard one which assumes a cosmological origin of the MBR.

However no such difficulty arises for $2/3\leq \Omega_{\rm m 0}\leq 1$
(or $0\leq (\Omega_{\rm m 0}-2\Omega_{\Lambda 0})\leq 1$, in the non-flat models),
in which case the model is either decelerating ($\Omega_{\rm m 0}>2/3$) or
coasting ($\Omega_{\rm m 0}=2/3$) throughout this phase of evolution.

We also note that $\rho_{\rm v}\frac{<}{>}\rho_{\rm m}$ according as
$z \frac{>}{<}(2\Omega_{\rm m 0}-1)/(2-3\Omega_{\rm m 0})$ if $\Omega_{\rm m 0}>
2/3$.
For $\Omega_{\rm m 0}<2/3$, $\rho_{\rm v}\frac{<}{>}\rho_{\rm m}$ according as
$z \frac{<}{>}(2\Omega_{\rm m 0}-1)/(2-3\Omega_{\rm m 0})$.
Thus, we find that with a positive $\Lambda$, the dependence of the relative
dominance of $\rho_{\rm v}$ and $\rho_{\rm m}$ on $z$ is just opposite to what
we have seen in the {\it quintessence} model. For example, in the model
$\Omega_{\rm m 0}=0.65$, $\rho_{\rm m}$ started dominating over $\rho_{\rm v}$
at $z=6$. It should not be surprising that $\rho_{\rm v}$ is decreasing with
time faster than $\rho_{\rm m}$ in this model. This is because of a continuous
creation of matter from the decaying $\rho_{\rm v}$, as we shall discuss later.
Equation (13), which implies that
$(\rho_{\rm m} -2 \rho_{\rm v}) \sim (1+z)^3$, explains why $\Omega_{\rm m}$
and $\Omega_{\rm v}$ are of the same order at the present epoch.
(Note that in the coasting model, $\Omega_{\rm m}=2\Omega_{\rm v}$ always).

Our fitting procedure gives the best-fitting solution as $\Omega_{\rm m 0}=0.59$
with $\chi^2=66.89$ (at 53 dof). This represents a reasonably good fit
($\Delta\chi^2=1.26$), though not as good as we have obtained for a
constant $\Lambda$. However, we note that this model can be rejected on the
grounds that it cannot accommodate objects with redshift higher than 2.57. But
the parameter space, allowed by the data, is wide enough to accommodate models
with high enough $z_{\rm max}$. For example, the model $\Omega_{\rm m 0}=0.65$
gives  $\chi^2=68.12$ (at 54 dof), which represents a
reasonably good fit ($\Delta\chi^2=1.26$).

Note that the best-fitting solution without including SN 1997ff is given by
$\Omega_{\rm m 0}=0.49$ with $\chi^2=58.98$ (at 52 dof, i.e.,
$\Delta\chi^2=1.13$). This
implies that the addition of SN 1997ff has worsened the fit considerably.

It is important to note that the quintessence fields, which acquire negative
pressure during the matter dominated phase and behave like  dynamical $\Lambda$,
seem formally
equivalent to the kinematical models of $\Lambda$ considered in this section.
However, these two categories are, in general, fundamentally different from
the theoretical point of view. In the former case, quintessence and matter
fields are assumed to be conserved separately (through the assumption of
minimal coupling of the scalar field with the matter fields). However, in the
latter case, the conserved
quantity is $[T^{ij}_{\rm m}-\{\Lambda(t)/8\pi G\}g^{ij}]$, leading to

\begin{equation}
\dot\rho_{\rm tot}+3(\rho_{\rm tot}+p_{\rm tot})\frac{\dot S}{S}=0,
\end{equation}
which follows from the Bianchi identities. (Obviously, the additional
requirement of the conservation of the matter field forces $\Lambda$ to become
a constant). This implies that there is a continuous creation of matter
from the {\it decaying} $\Lambda$ as is clear from the following

\begin{equation}
\rho_{\rm m}=C S^{-3(1+\omega)}-\frac{S^{-3(1+\omega)}}{8\pi G}
\int \dot\Lambda(t) S^{3(1+\omega)}{\rm d}t, ~ ~ ~ C=\mbox{constant},
\end{equation}
which follows from (17). Thus with a positive kinematical $\Lambda (t)$, the
energy momentum tensor of the $\Lambda$-field $\equiv -(\Lambda/8\pi G)g^{ij}$
(with its equation of state $p_\Lambda=-\rho_\Lambda=-\Lambda/8\pi G$) acts
just like the creation field of the Quasi-Steady State Cosmology
(which will be discussed in section 6). Because
of its negative energy, the $c-$field also has a repulsive effect to accelerate
the expansion as does a positive $\Lambda$.

Due to this fundamental difference, the two categories of the models have,
in general, completely different dynamics of evolution. Note that
$\rho_{\rm m}$, in equations (13) and (19), no longer scales down as
$(1+z)^{3(1+\omega)}$ as it does in equation (5).
In this connection, we consider an interesting example of the
quintessential model with $\omega_\phi=-1/3$ (which is
the equation of state of the non-relativistic cosmic string). Note that the
resulting model also has an
$S^{-2}$-variation of the effective $\Lambda$, i.e.,
$\Lambda\equiv 8\pi G \rho_\phi\sim S^{-2}$. However, the evolutions
of $\rho_{\rm m}$ and $H$ obtained from equations (5) and (11), in this flat model,
are completely different from those in (13) and (16). This is also obvious
from figure 1, where we have shown the expansion dynamics of these models,
together with some other quintessential and kinematical $\Lambda$-models and compared
them with the constant $\Lambda$-model.

\vspace{.7cm}

\centerline{{\epsfxsize=14cm {\epsfbox[50 250 550 550]{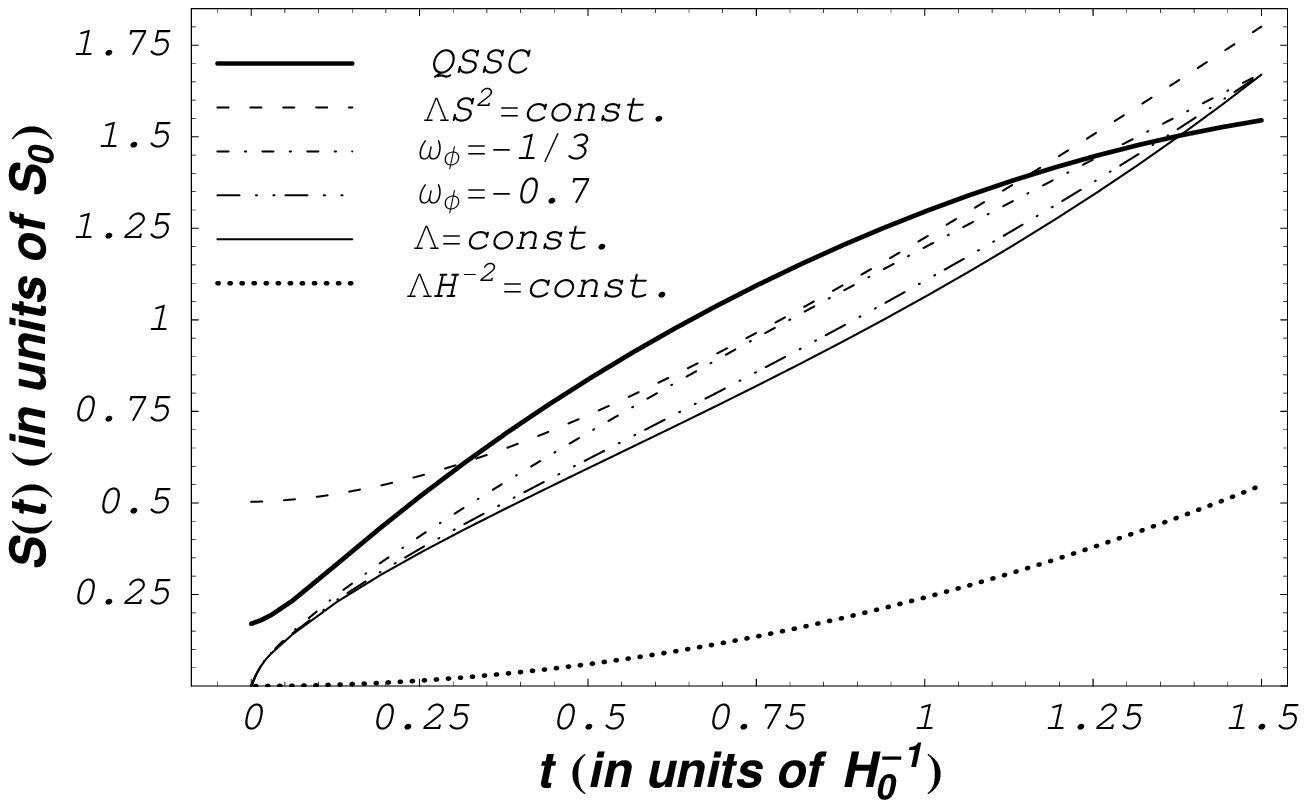}}}}

{\bf Figure 1:} The time-evolution of the scale factor $S$ (in units of $S_0$) is

shown as a function of the cosmic time $t$ (in units of $H_0^{-1}$), in some flat

models. The plotted quasi-steady state model is given by $\Omega_{\Lambda0}=-0.358$

with $z_{\rm max}=5$, whereas the rest of the models are specified by
$\Omega_{\rm m0}=0.33$.

\vspace{1cm}

\noindent
{\bf Case (2): $\Lambda =n_2 H^2$}

\vspace{.2cm}
\noindent
The (flat) model is described by the following equations:

\begin{equation}
\rho_{\rm m}=C_2 S^{(n_2-3)(1+w)}, ~  ~
~ C_2=\mbox{constant}>0,
\end{equation}

\begin{equation}
\dot{S}^2=\frac{8\pi G C_2}{(3-n_2)} S^{(n_2-3)(1+w)+2}.
\end{equation}
Note that the ansatz $\Lambda=n_2 H^2$ is equivalent to assuming that
$\Omega_{\Lambda}$ $(=n_2/3)$ is a constant and, hence, so is $\Omega_{\rm m}$
$(=1-n_2/3)$ in a flat model. We write $n_2$, and also $C_2$, in terms of the
parameters at the present epoch as

\begin{equation}
n_2=3(1-\Omega_{\rm m0}), ~ ~ ~ C_2=\frac{3H_0^2}{8\pi G}\Omega_{\rm m0}
S_0^{3\Omega_{\rm m0}}.
\end{equation}
The expression for the Hubble parameter, in the present epoch, then yields

\begin{equation}
H(z)=H_0(1+z)^{3\Omega_{\rm m0}/2}.
\end{equation}
We note that both, $\rho_{\rm m}$ and $\rho_{\rm v}$ scale down as
$(1+z)^{(3-n_2)(1+\omega)}$ maintaining a fixed ratio
$\rho_{\rm m}/\rho_{\rm v}=\Omega_{\rm m}/(1-\Omega_{\rm m})$.
The deceleration parameter, in the model, is obtained as

\begin{equation}
q=\frac{(3-n_2)(1+\omega)}{2}-1,
\end{equation}
which is constant and implies that $q\frac{>}{<}0$ according as
$n_2\frac{<}{>}(1+3\omega)/(1+\omega)$. Thus two different values of the parameter
$n_2$, viz., one with $n_2<3/2$ in the early radiation dominated era and the other
with  $n_2>1$ in the present era can make the universe shift from deceleration
to acceleration. In this view, $n_2$ behaves like a {\it tracking} parameter.

The fitting of the 54 points-data to the model gives the best-fitting solution
as $\Omega_{\rm m0}=0.4$ with $\chi^2=58.35$ (at 52 dof), which represents a good
fit ($\Delta \chi^2=1.12$). The addition of
SN 1997ff to this dataset worsens the fit and the new dataset gives the
best-fitting solution as $\Omega_{\rm m0}=0.49$ with $\chi^2=65.27$ (at 53 dof,
i.e., $\Delta \chi^2=1.23$), which though
represents a reasonably good fit, better than the one we obtained for the model
$\Lambda\sim S^{-2}$, but not as good as the ones obtained for the constant
$\Lambda$-models.

From the best-fitting solutions of the models discussed so far (see, Table 1),
we notice that
the kinematial $\Lambda$-models require lesser amount of {\it dark energy}
($\Omega_{\Lambda 0}$) than is
required by the quintessential or constant $\Lambda$-models, to fit the data.
Said in other words, this means that with a given amount of {\it dark energy} density,
the kinematial $\Lambda$-models show larger distance of an object of a
given redshift than the distance the object has in  the quintessential or
constant $\Lambda$-models. This is also clear from figure 2, where we have
compared the {\it luminosity distance} of a source (at different redshifts $z$)
calculated in different models.

\newpage

\centerline{{\epsfxsize=14cm {\epsfbox[50 250 550 550]{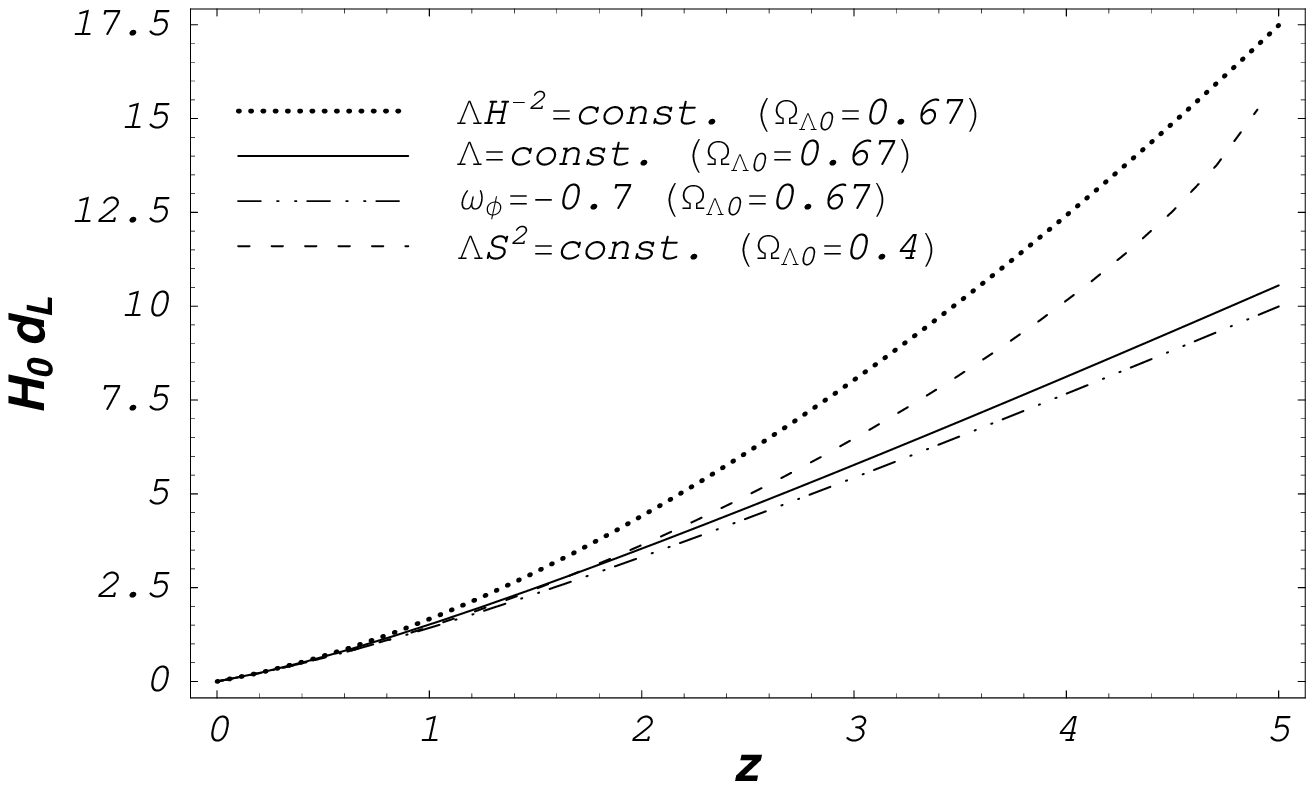}}}}

{\bf Figure 2:} The {\it luminosity distance} $d_{\rm L}$ (in units of $H_0^{-1}$)
is plotted as a

function of the cosmological redshift in some flat models. With a given

amount of {\it dark energy} density, the kinematial $\Lambda$-models show larger

distance of an object at a given redshift than the distance the object has

in the quintessential or constant $\Lambda$-models. The model $\Lambda\sim S^{-2}$,
with

even a smaller $\Omega_{\Lambda0}=0.4$, shows higher distance
of a source than it has

in the constant $\Lambda$- or the quintessential model with $\Omega_{\Lambda0}=0.67$. Note

that all these models are specified completely by only one parameter $\Omega_{\Lambda0}$.

\vspace{1.5cm}
It seems that all the best-fitting models discussed so far
fit the data reasonably well. In order to be able to discriminate between
the models, we magnify their differences by plotting the relative magnitude with
respect to a fiducial constant $\Lambda$-model $\Omega_{\rm m0}=0.3$ and $\Omega_\Lambda0=0.7$,
as has been shown in figure 3. However,
we realize that it is still not possible to distinguish the
models with the available accuracy of the data. Some more high redshift supernovae
with $z>1$
are needed to make the discrimination possible, as the models start to depart
from each other very rapidly once we go above $z=1$, as is clear from figure 4.

\vspace{.7cm}

\centerline{{\epsfxsize=14cm {\epsfbox[50 250 550 550]{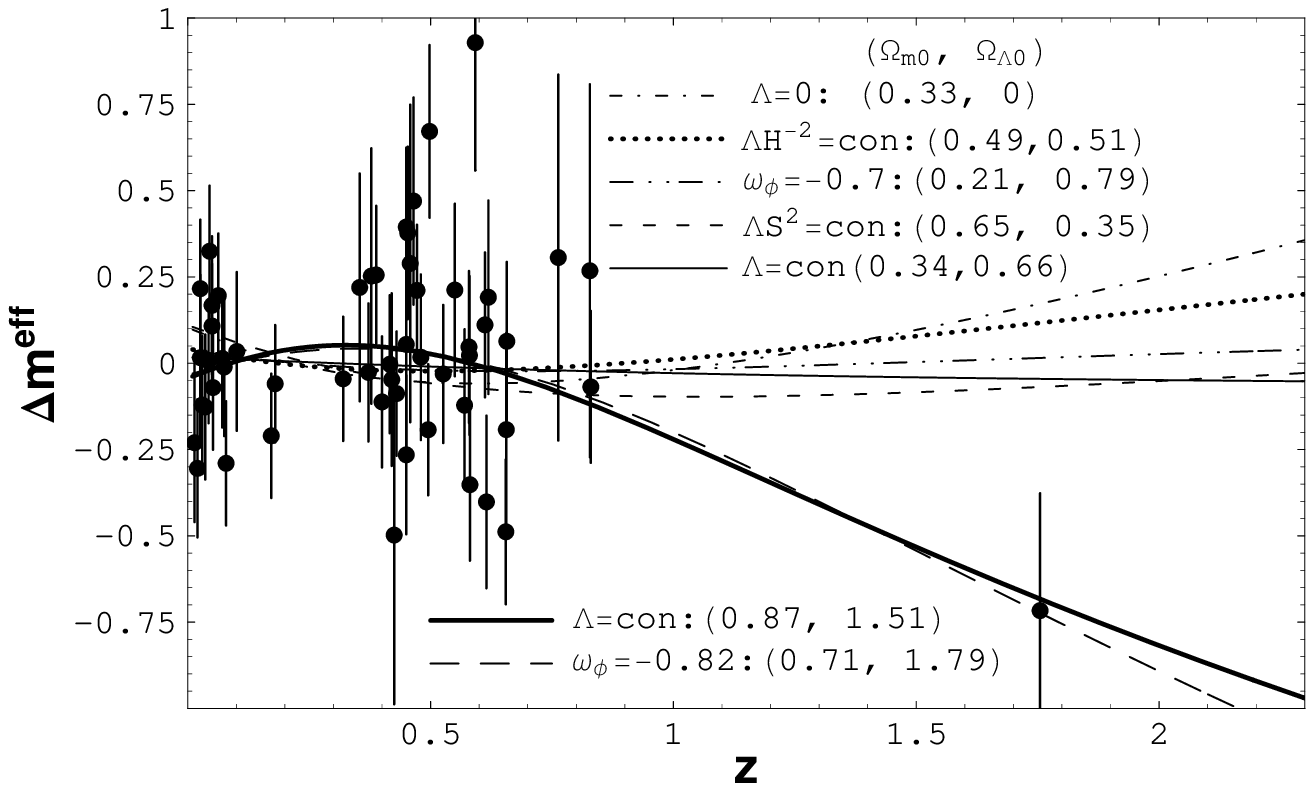}}}}

{\bf Figure 3:} Hubble diagram of the SN Ia minus a fiducial flat FRW model

with $\Omega_{\rm m0}=0.3$ (i.e., the relative magnitude
$\Delta m(z)\equiv m(z)-m_{{\rm fiducial}}(z)$):

The curves correspond to some best-fitting theoretical models. The older

data points ($z<1$) have too large error bars to make a discrimination

possible between the models (especially between the flat models including

the open model with zero $\Lambda$ and whisker-dust). However, the new point

(SN 1997ff) seems to be consistent with only the best-fitting spherical

models.

\vspace{1cm}

\centerline{{\epsfxsize=14cm {\epsfbox[50 250 550 550]{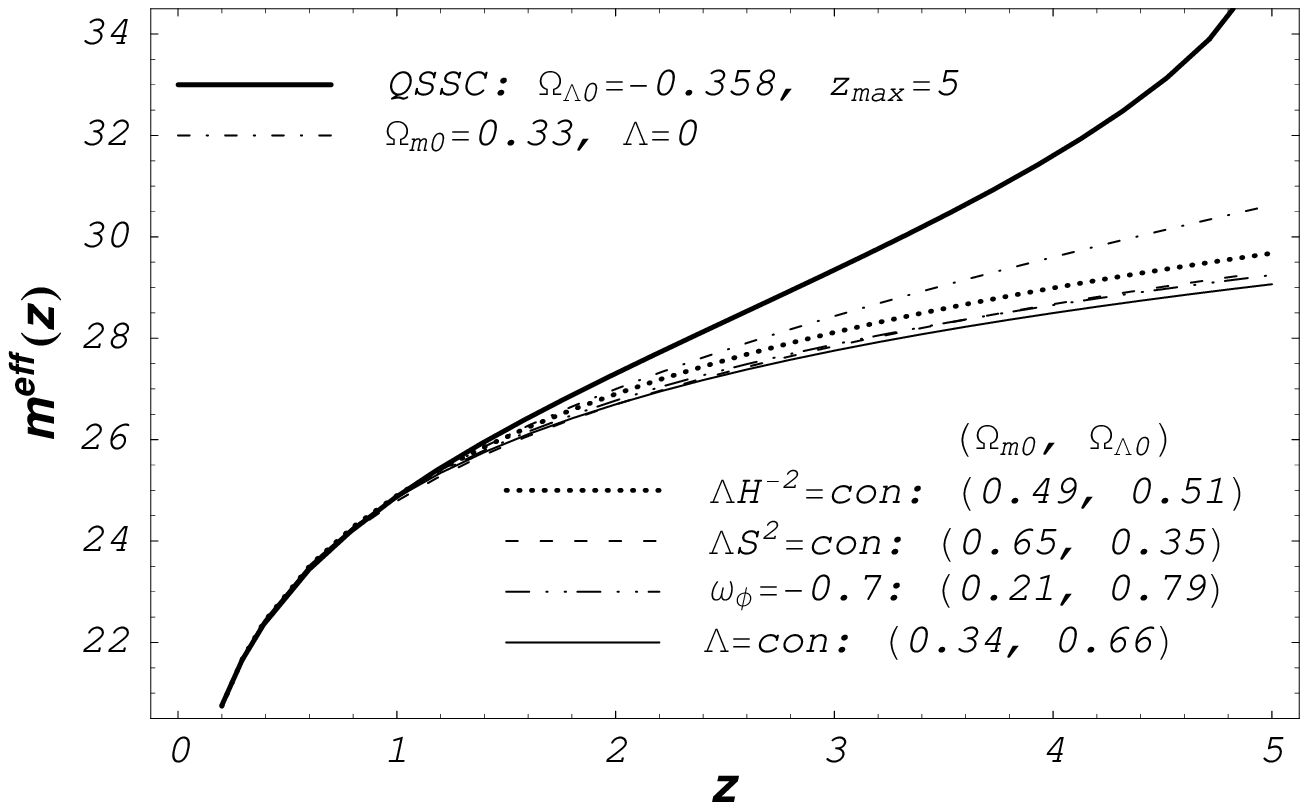}}}}

{\bf Figure 4:} The effective magnitude is shown as a function of the redshift

in some best-fitting  flat models and the open FRW model with
zero $\Lambda$.

The curves start to depart from each other very rapidly as we go above

$z=1$.

\vspace{1cm}

\noindent
{\bf 4. EFFECT OF THE INTERGALACTIC DUST}

\vspace{.2cm}
\noindent
It has been argued that the dimming of the high redshift supernovae relative to
the low redshift ones could be an artefact of the intervening dust
(Aguire, 1999; Banerjee et al,
2000). The intergalactic dust
is claimed to be dominated by needle- or whisker-type dust which is supposed to
form from the condensation of the metallic vapours ejected from the supernovae
explosions (Hoyle \& Wickramasinghe, 1988; Narlikar et al, 1997).
It is confirmed from the laboratory experiments that this needle-like metallic
dust can grow more rapidly than the spherical dust (Hoyle et al, 2000).

In a homogeneous, isotropic universe, the contribution to the effective magnitude
arising from the
absorption of light by the intervening whisker-like dust, is given by

\begin{equation}
\delta m(z)=\int_0^{\ell(z)} \kappa \rho_g {\rm d}\ell=\kappa \rho_{g0}\int_0^z
(1+z')^2\frac{{\rm d}z'}{H(z')},
\end{equation}
where $\kappa$ is the mass absorption coefficient, which is effectively
constant over a wide range of wavelengths and is of the order $10^5$
cm$^2$ g$^{-1}$ (Wickramasinghe \& Wallis, 1996),
$\rho_g\sim S^{-3}$ is the whisker grain density with
$\rho_{g0}$ in the range $(3 - 5)\times 10^{-34}$
g cm$^{-3}$ and $\ell(z)$ is the proper distance traversed by light through
the intergalactic medium emitted at the epoch of redshift $z$.

Thus the net magnitude is given by

\begin{equation}
 m^{\rm net}(z)=m(z) + \delta m(z),
\end{equation}
where the first term on the r.h.s. corresponds to the usual magnitude from the
cosmological evolution. We consider a particular FRW model with
$\Omega_{\rm m 0}=0.33$ and $\Lambda=0$ to test equation (25).
Using $H_0=70$ km s$^{-1}$ Mpc$^{-1}$, we find that the model fits the 54
point-data
(i.e., without including SN 1997ff) very well giving $\chi^2=59.78$ (at 52 dof),
which represents a good fit ($\Delta\chi^2=1.15$). However, the
addition of SN 1997ff to this data worsens the fit considerably and the
$\chi^2$ value increases to 70.38 (at 53 dof, i.e., $\Delta\chi^2=1.33$),
which though acceptable, is not as good as the ones with a non-zero $\Lambda$.

\vspace{1cm}

{\bf Table 1.} For a comparative study, the best-fitting parameters of the

models discussed in the paper are listed, which have been estimated from

the SN Ia data, with and without SN 1997ff. The parameters of all the

models, except the QSSC, are estimated from the SN sample of Permutter

et al' primary fit C (i.e., the excluded points are 1992bo, 1992br, 1994H,

1997O, 1996cg, 1996cn). In the case of the QSSC, the fitted data sample

is from Banerjee et al (where the excluded points are: 1992bc, 1992br,

1994H, 1997O, 1995as, 1995K). All the other models, except the first and

the last ones, shown in the table, are flat models.

\begin{center}
\begin{tabular}{lll}
\hline\hline
\multicolumn{1}{c}{Models} &
\multicolumn{1}{c}{with SN 1997ff}&
\multicolumn{1}{c}{without SN 1997ff}\\
 &\begin{tabular}{lll} $\Omega_{\rm m0}$& $\Omega_{\Lambda0}$ &   ~$\Delta 
 \chi^2$
 \end{tabular}&
\begin{tabular}{lll}$\Omega_{\rm m0}$  &$\Omega_{\Lambda0}$ & ~$\Delta \chi^2$ 
\end{tabular}\\
\hline

\vspace{.3cm}
$\omega_\phi=-1$ &\begin{tabular}{lll} 0.87 & 1.51 & 1.09 \end{tabular}&
\begin{tabular}{lll} 0.79 & 1.40 & 1.11  \end{tabular}\\

 &\begin{tabular}{ll} $\Omega_{\rm m0}^{\rm Flat}$  &  ~  $\Delta\chi^2$
 \end{tabular}&
\begin{tabular}{ll}$\Omega_{\rm m0}^{\rm Flat}$  &  ~ $\Delta\chi^2$
\end{tabular}\\
\hline

$\omega_\phi=-1$& \begin{tabular}{ll} 0.34 & ~ 1.17  \end{tabular}&
\begin{tabular}{ll} 0.28 & ~ 1.11  \end{tabular}\\

$\omega_\phi=-0.9$& \begin{tabular}{ll} 0.30 & ~ 1.18  \end{tabular}&
\begin{tabular}{ll} 0.24 & ~ 1.11  \end{tabular}\\

$\omega_\phi=-0.8$& \begin{tabular}{ll} 0.26 & ~ 1.19  \end{tabular}&
\begin{tabular}{ll} 0.18 & ~ 1.11 \end{tabular}\\

$\omega_\phi=-0.7$& \begin{tabular}{ll}0.21  & ~ 1.20  \end{tabular}&
\begin{tabular}{ll} 0.11 & ~ 1.12  \end{tabular}\\

$\omega_\phi=-0.6$& \begin{tabular}{ll} 0.14 & ~ 1.21  \end{tabular}&
\begin{tabular}{ll} 0.01 & ~ 1.12 \end{tabular}\\

$\Lambda \sim H^2$ & \begin{tabular}{ll} 0.49 & ~ 1.23  \end{tabular}&
\begin{tabular}{ll}0.40  & ~ 1.12 \end{tabular}\\

 \vspace{.3cm}
$\Lambda \sim S^{-2}$ &\begin{tabular}{ll} 0.59 & ~ 1.26  \end{tabular}&
\begin{tabular}{ll}0.49  & ~ 1.13 \end{tabular}\\

\vspace{.3cm}
\begin{tabular}{l}QSSC: $z_{\rm max}=5$,\\$\Omega_{\Lambda0}=-0.358$ \end{tabular}
& \begin{tabular}{ll} ~  ~ ~ & ~ 1.27  \end{tabular}&
\begin{tabular}{ll} ~ ~ ~~& ~ 1.00 \end{tabular}\\

\begin{tabular}{l}FRW: $\Lambda=0$,\\$\Omega_{\rm m0}=0.33$ \end{tabular}
& \begin{tabular}{ll} ~  ~ ~ & ~ 1.33 \end{tabular}&
\begin{tabular}{ll} ~ ~ ~~& ~ 1.15  \end{tabular}\\
\hline
\end{tabular}
\end{center}

\vspace{1cm}

\noindent
{\bf 5. AGE CONSIDERATIONS}

\vspace{.2cm}
\noindent
The age of the universe $t_0$, in a model with the Hubble parameter $H(z)$,
is given by

\begin{equation}
t_0=\int_0^{z_{\rm max}}\frac{H^{-1}(z)}{(1+z)}{\rm d} z.
\end{equation}
The value of $t_0$ depends on the expansion dynamics of the model considered and, hence, on the
free parameters of the model. Additionally, the age of the universe depends on
the present value of the Hubble parameter $H_0$. We consider
$H_0=70$ km s$^{-1}$ Mpc$^{-1}$ which comes from the present consensus
of the Hubble Constant proponents (Freedman et al, 2001).
In figure 5, we have plotted $t_0$ as a function
of $\Omega_{\Lambda0}$ in the models discussed in the preceding sections.
We note that the age of the universe in the model with a constant $\Lambda$ is
uncomfortably close to the age of the globular clusters $t_{\rm GC}=12.5$ Gyr,
even for
$\Omega_{\rm m0}$ as low as 0.33. The age in the {\it quintessential} models is
even lower. Note that if the required mass density $\Omega_{\rm m0}$ of the universe
was smaller, one could get higher age in these models, as is clear from the
figure. However, this does not seem likely as the recent measurements give very
narrow range of $\Omega_{\rm m0}$ as $\Omega_{\rm m0}=0.330\pm 0.035$ at one sigma level (Turner, 2001).
We also note that there are various other sources of supporting evidence that are
consistent with this value, including studies of the evolution of cluster
abundances with redshift, measurements of the power spectrum of large-scale
structure, analyses of measured peculiar velocities as they relate to the observed
matter distribution, and observations of the outflow of material from voids.
If $\Omega_{\rm m0}$ was, on the other hand, larger than this value, one could perhaps argue
that there was some matter in an yet undetected form.

In a flat model with a constant $\Lambda$, the fractional uncertainty in the value
of $t_0$ can be calculated, from equations (11) and (26), as

\begin{equation}
\frac{\Delta t_0}{t_0}=\frac{\Delta H_0}{H_0} +
\frac{\int_0^1 [\Omega_{\rm m0} x^{-1}+(1-\Omega_{\rm m0})x^2]^{-3/2}
(x^{-1}-x^2) {\rm d}x}
{2 \int_0^1 [\Omega_{\rm m0} x^{-1}+(1-\Omega_{\rm m0})x^2]^{-1/2}
{\rm d}x}\Delta \Omega_{\rm m0}.
\end{equation}
The recent measurement of $H_0$, by using {\it Hubble Space Telescope}, gives the
fractional error in $H_0$ as $\approx$10 percent (Freedman et al, 2001).
Thus the expected error in $t_0$ in this model with $\Omega_{\rm m0}=0.33$
comes out as about 13 percent (i.e., $t_0 =13.1$ Gyr, $\Delta t_0\approx 1.7$ Gyr).
On the other hand, the age of the oldest globular cluster is estimated to be
$t_{\rm GC}=12.5\pm 1.2$ Gyr (Gnedin et al, 2001; and the references therein).
Additionally, the latest uranium decay estimates yield the age of the Milky Way
as $12.5\pm 3$ Gyr (Cayrel et al, 2001). Thus, if one compares the upper age limit
of these objects with the lower limit of the age of the universe in
this model, the model is in serious trouble. This situation in
the {\it quintessential} models is even worse, as is clear from the figure.
The situation with the best-fitting closed
model with constant $\Lambda$ ($\Omega_{\rm m0}=0.79$, $\Omega_{\Lambda 0}=1.41$)
is not better either, which also gives $t_0 =13.1$ Gyr. The open model
$\Omega_{\rm m0}=0.33$, $\Lambda=0$ yields $t_0 =11.2$ Gyr only.
However, the kinematical $\Lambda$-models do not face this problem,
{\it especially, the model $\Lambda\sim H^2$, which gives a remarkably large
value of $t_0=19$} Gyr {\it in the best-fitting model}. For a lower value
of $\Omega_{\rm m0}$,  $t_0$ is even higher. For example, the value
$\Omega_{\rm m0}=0.33$ in this model gives $t_0 =28.2$ Gyr!
This is not surprising
because the expression for the age of the universe in this model yields
\begin{equation}
t_0 =\frac{2}{3\Omega_{\rm m0}}H_0^{-1},
\end{equation}
which gives higher values of $t_0$ for lower $\Omega_{\rm m0}$. The model
$\Lambda\sim S^{-2}$ (with even a higher $\Omega_{\rm m0}=0.65$), also gives a
comfortable value $t_0=14.7$ Gyr.

\vspace{1cm}

\centerline{{\epsfxsize=14cm {\epsfbox[50 250 550 550]{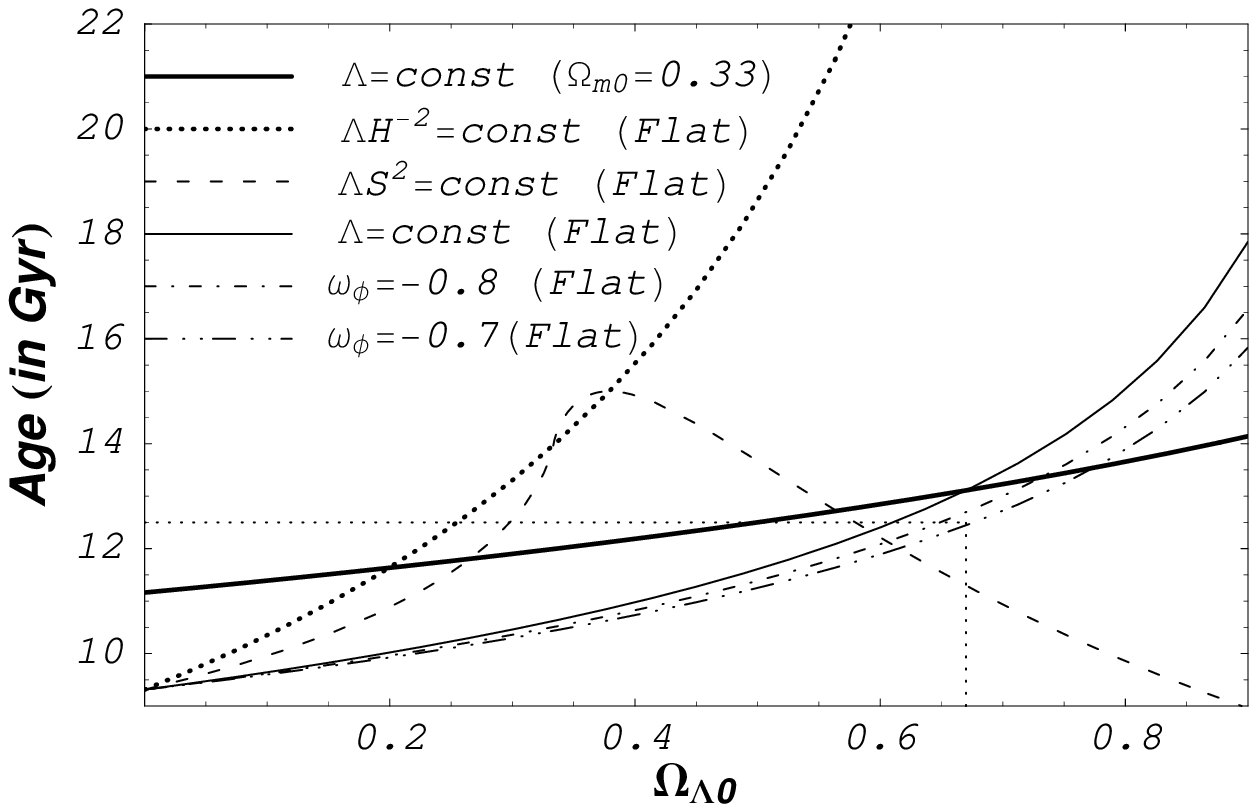}}}}

{\bf Figure 5:} The age of the universe is plotted as a function of
$\Omega_{\Lambda0}$ in

different models, by using $H_0=70$ km s$^{-1}$ Mpc$^{-1}$. The horizontal dotted

line represents the age of the globular clusters $t_{\rm GC}=12.5$ Gyr. The

vertical dotted line corresponds to the mass density of the present universe

($\Omega_{\rm m0}=0.33$), in the flat models ($\Omega_{\rm tot}=1$). In the
non-flat constant

$\Lambda$-model, $\Omega_{\rm tot}=0.33+\Omega_{\Lambda0}$.

\vspace{1cm}
\noindent
{\bf 6. QUASI-STEADY STATE COSMOLOGY}

\vspace{.2cm}
\noindent
Another cosmological theory, where the $\Lambda$-{\it energy} plays an important
role is the Quasi-Steady State Cosmology (QSSC) which is a Machian theory proposed
by Hoyle, Burbidge and Narlikar as an alternative to the Standard Big Bang
Cosmology (SBBC). A built-in negative $\Lambda$ in this theory, which is not
primarily from the observations, is necessary to make the solutions possible.
Its role in the dynamics of the model is to energize the creation-field by controlling the
expansion of the universe. The repulsive effects, akin to that from a positive
$\Lambda$, are produced by the creation field which has a negative energy density.

This cosmology does not have any cosmic epoch when the universe was hot and {\it
is free from the initial singularity} of the SBBC. It has cycles of expansion and
contraction (regulated by the creation- and the negative $\Lambda$-fields
respectively) of comparatively shorter period superposed on a long term
exponential trend of expansion. Creation of matter is also periodic, being
confined to pockets of strong gravitational fields around compact massive objects.
The theory seems to meet all the available observational constraints at the
present time. According to QSSC, the MBR is the relic starlight left by the stars
of the previous cycles which has been thermalized by the whisker dust emitted
by the supernovae. It is very interesting to note that {\it the energy available
from
this process is just right to give a radiation background of 2.7 K at the present
epoch} (Hoyle et al, 1994). SBBC, on the other hand, does not predict the present
temperature of the MBR. Another point in favour of the QSSC is that the model
(which, though, has a constant $\Lambda$) does not face the cosmological constant
problem mentioned in section 1 and the magnitude of $\Lambda$ is broadly related
with the size of the universe in a Machian way (Hoyle et al, 1995).
In fact, the $\Lambda$ in the QSSC does not represent
the energy density of the quantum fields, as this model does not experience
the energy scales of particle physics.

In the following, we describe briefly the dynamics of the model. (For a detailed
discussion, see Sachs et al, 1996; and the references therein).
The field equations of QSSC, in effect, are the Einstein field equations together
with a (negative) $\Lambda$ and a trace-free zero rest mass scalar field $c$
representing the creation of matter:

\begin{equation}
R^{ij}-\frac{1}{2} R g^{ij}=-8\pi G\left[T^{ij}-
\frac{\Lambda (t)}{8\pi G} ~ g^{ij} -f\left(c^ic^j+\frac{1}{4}c^\ell c_\ell g^{ij}
\right)\right],
\end{equation}
where $f$ is a coupling constant of the $c$-field to gravity and
$c_i\equiv \partial c/\partial x^i$. Assuming that the present epoch is represented
by the non-creative mode of the model (i.e., $T^{ij}_{{\rm (m)};j}=0$) and the
matter is in the form of dust, equations (29), in the background of R-W metric,
yield the following three independent equations:

\begin{equation}
\rho_{\rm m} \sim (1+z)^3, ~ ~ \dot c \sim (1+z)^2,
\end{equation}

\begin{equation}
H(z)=H_0[\Omega_{\Lambda0}-\Omega_{k0}(1+z)^2+\Omega_{\rm m0}(1+z)^3+
\Omega_{c0}(1+z)^4]^{1/2},
\end{equation}

\begin{equation}
2 q(z)= \left[\frac{H_0}{H(z)}\right]^2[\Omega_{\rm m0}(1+z)^3-2\Omega_{\Lambda0}
+2\Omega_{c0}(1+z)^4],
\end{equation}
where $\Omega_c\equiv 8\pi G\rho_c/3H^2$, with $\rho_c\equiv-3f\dot c^2/4$
and $p_c\equiv-f\dot c^2/4$.
For $k=0$, equation (31) reduces to

\begin{equation}
S=\bar S[1+\eta \cos \psi(t)],
\end{equation}
where $\eta$ is a parameter satisfying $0<\eta<1$, $\bar S$ is a constant given
in terms of the constants  $\Omega_{i0}$'s and $\eta$ and the function $\psi$ is
given by

\begin{equation}
\dot\psi^2=-\frac{\Lambda}{3}(1+\eta\cos\psi)^{-2}\{6+4\eta\cos\psi+\eta^2(1+
\cos^2\psi)\}.
\end{equation}
Clearly, $\Lambda$ has to be negative. Thus the typical {\it energy} represented
by the  $\Lambda$ in QSSC (let us call it {\it $\Lambda$-energy}) has a negative
energy density ($=-\mid\Lambda\mid/8\pi G$) and a positive pressure
($=\mid\Lambda\mid/8\pi G$).
It is obvious from equation (33) that $S$ never becomes zero and oscillates
between
\begin{equation}
\bar S(1-\eta)\equiv S_{\rm min}\leq S\leq S_{\rm max} \equiv \bar S(1+\eta).
\end{equation}
Thus at the maximum redshift $z_{\rm max}$ (in the present cycle, say), we have the
following identity from equation (31):

\begin{equation}
\Omega_{\Lambda0}-\Omega_{k0}(1+z_{\rm max})^2+\Omega_{\rm m0}(1+z_{\rm max})^3+
\Omega_{c0}(1+z_{\rm max})^4=0.
\end{equation}

As the QSSC assumes the abundance of whisker-like dust present in the intergalactic
medium which is essential to explain the MBR, the $m-z$ relation in the model is
given by equation (25). Banerjee et al (2000) have already studied the fitting
of the SN Ia data to this model by considering a sample of 54 data points which
was obtained by excluding those 6 points from the full sample of
60 points, {\it for which the difference between the theory and the observation
was excessive}. By assuming $z_{\rm max}=5$, they considered a particular QSS
model
$\Omega_{\Lambda0}=-0.358$ for the actual fitting to this sample. The model shows
an excellent fit to the sample giving $\chi^2=51.97$ (at 52 dof, i.e.,
$\Delta\chi^2=1.00$). The particular values of the
parameters come from the constraints from various observations, though it is not
essential to assume only these parameters. Indeed, the parameter space,
which fits the data, is wide enough which makes the model robust.
When we add the new point SN 1997ff to this sample, the model still fits the
resulting dataset
reasonably well for a wide range of parameters, as is clear from figure 6, where
we have shown a parameter-space $0>\Omega_{\Lambda0}\geq -0.32$ which has the
$\Delta\chi^2$ value in the range $1.16 - 1.26$.
If one sticks to the model $\Omega_{\Lambda0}=-0.358$, then the value of
$\chi^2$ increases to 67.41 (at 53 dof, i.e., $\Delta\chi^2=1.27$), which
is also acceptable.
\vspace{1cm}

\centerline{{\epsfxsize=14cm {\epsfbox[50 250 550 550]{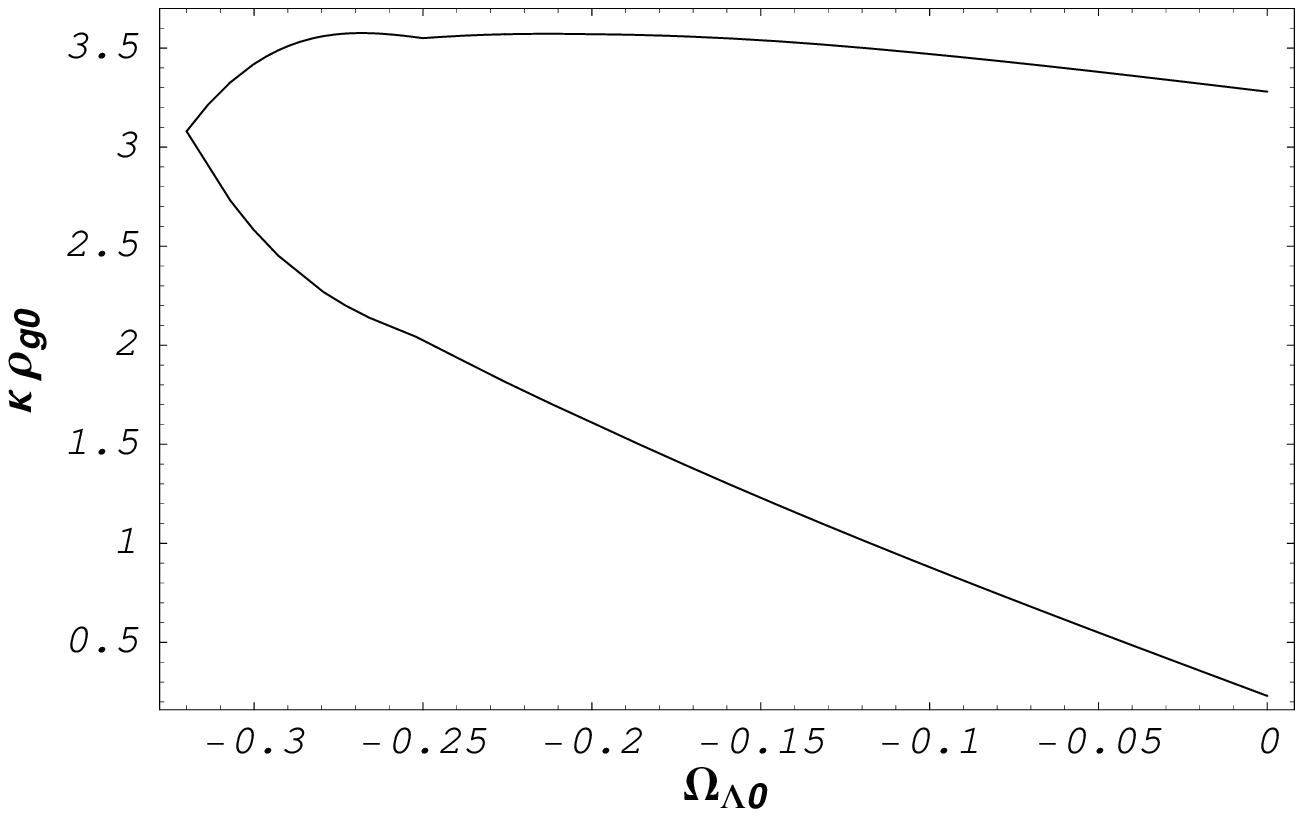}}}}

{\bf Figure 6:} A parameter-space of interest in the QSSC: the points of the

contour have $\Delta\chi^2$ in the range $1.16 - 1.26$.
The coupling constant $\kappa$ and

the present density of the whisker grains $\rho_{g0}$
are measured, respectively,

in units of $10^5$ cm$^2$ g$^{-1}$ and $10^{-34}$ g cm$^{-3}$.

\vspace{1cm}
\noindent
CONCLUSION

\vspace{.2cm}
\noindent
We have studied two kinds of popular explanations of the modified SN data
(Perlmutter st al' data together with
the recently observed SN 1997ff): the accelerating universe-interpretation in some
{\it dark energy} models of variable  and constant $\Lambda$; and the absorption
of light by the intergalactic dust in the standard big bang model and the
quasi-steady state model.
The addition of the new point SN 1997ff to the old data has worsened
the fit to most of the models, except a closed FRW model with a constant $\Lambda$,
and a closed quintessential model with $\omega_\phi=-0.82$,
where it has improved the fit. In the flat quintessential models, the minimum value
of $\chi^2$ decreases as the equation of state-parameter $\omega_\phi$ decreases.
Thus, under the constraint of the {\it null energy condition}, the best-fitting
quintessential model has $\omega_\phi$ close to $-1$.
The kinenatical $\Lambda$-models $\Lambda\sim S^{-2}$ and $\Lambda\sim H^2$,
considered in the paper, also fit the data reasonably well, though the fit is not
as good as in the constant $\Lambda$-case. However, the model $\Lambda\sim S^{-2}$
with low energy density becomes unphysical as it cannot accommodate higher redshift
objects. It is found that the kinenatical $\Lambda$-models require more matter
energy density to fit the data than is required by the constant $\Lambda$-model.
 We find that the expansion of the
universe shifted from deceleration to acceleration at about $z=0.57$, if the
universe is flat and the dark energy is represented by a constant $\Lambda$.

When we examine the alternative explanation of the data by considering the
absorption of radiation by the intervening whisker-like dust, we find that the
quasi-steady state model and the FRW model $\Omega_{\rm m0}=0.33$, with $\Lambda=0$
also fit the data reasonably well, though
the fit is not as good as the one to the constant $\Lambda$-model.

While the older data was showing an almost same goodness of fit to all the models
(see Table 1), the addition of the new point has deteriorated the fit to most of
the models considerably and,
if taken on their face values, the modified data show a tendency to favour
a closed universe with the dark energy in the form of either
a pure cosmological constant or a {\it quintessence} with $\omega_\phi=-0.82$.
However, the fit of the data to other models is also reasonably good and
some more points with $z>1$ and smaller error bars would be needed to
enable a clear discrimination between the different models. This might be accomplished
by a dedicated space telescope {\it SuperNova Acceleration Probe} (SNAP)
(Weller \& Albrecht, 2001) which aims to give the accurate luminosity distance
of $\approx$2000 type Ia supernovae up to $z\approx 1.7$ every year.

We have also studied the age of the universe in these models and find that the
models with a constant $\Lambda$ show an age which is uncomfortably close to
the age of the globular clusters. The quintessential models yield even lower age.
The kinematical $\Lambda$-models are, however, interesting in this view, which
give
large enough age, especially the model $\Lambda\sim H^2$ which yields a remarkably
large age of 28.2 Gyr for the current estimate of $\Omega_{\rm m0}=0.33$.
Note that in the QSSC, there is no beginning of the universe.

\vspace{1cm}

\noindent
{\bf ACKNOWLEDGEMENTS}

\noindent
The author thanks Shiv Sethi and Tarun Deep Saini for discussion and J. V.
Narlikar for discussion and careful reading of the draft and making valuable
comments. The author is grateful to the Department of Atomic energy for providing
the Homi Bhabha postdoctoral fellowship and IUCAA for hospitality. Thanks are
also due to Adam Riess for sending his comments and data and to
an anonymous referee for useful comments.
\vspace{1.5cm}

\noindent
{\bf REFERENCES:}\\
Aguire A. N., 1999, ApJ, 512, L19\\
Avelino P. C., Martins C. J. A. P., astro-ph/0106274\\
Banerjee S. K., Narlikar J. V., Wickramasinghe N. C., Hoyle F., Burbidge,

\hspace{.5cm} G., 2000, ApJ, 119, 2583

\noindent
Bahcall N. A., Ostiker J. P., Perlmutter S., Steinhardt P. J., 1999, Science,
284, 1481\\
Berman M. S., Som M. M., 1990, Int. J. Theor. Phys., 29, 1411\\
Bertolami O., 1986, Nuovo Cimento B, 93, 36\\
Bertschinger E., 1998, ARA \&A, 36, 599\\
Boisseau J. R., Wheeler J. C., 1991, Astron. J., 101, 1281\\
Branch D. 1981, Astrophys. J., 248, 1076\\
Branch D. 1982, Astrophys. J. Lett., 316, L81\\
Carvalho J. C., Lima J. A. S., Waga I. 1992, Phys. Rev. D, 46, 2404\\
Cayrel R., et al, 2001, Nature, 409, 691\\
Chen W., Wu Y. S. 1990, Phys. Rev. D, 41, 695\\
Dodelson S., Knox L., 2000, Phys. Rev. Lett., 84, 3523\\
Freedman W. L. 2001, Astrophys. J. 553, 47\\
Gnedin O. Y., Lahav O., Rees M. J., astro-ph/0108034 \\
Hoyle F., Wickramasinghe N. C., 1988, Astrophys. Space Sc. 147, 245\\
Hoyle F., Burbidge G., Narlikar J. V., 1994, MNRAS, 267, 1007\\
Hoyle F., Burbidge G., Narlikar J. V., 1995, Proc. R. Soc. London A, 448,

\hspace{.5cm}  191

\noindent
Hoyle F., Burbidge G., Narlikar J. V., 2000, {\it A Different Approach to

\hspace{.5cm}       Cosmology}, (Cambridge: Cambridge Univ. Press)

\noindent
Jaffe et al 2001, Phys. Rev. Lett. 86, 3475 \\
Kofman L. A., Gnedin N. Y., Bahcall N. A., 1993, Astrophys. J., 413, 1\\
Krauss L., Turner M. S., 1995, Gen. Rel. Grav., 27, 1135\\
Lau Y. K., 1985, Aust. J. Phys., 38, 547\\
Lewis G. F., Ibata R. A.  astro-ph/0104254 \\
Lopez J. L., Nanopoulos D. V., 1996, Mod. Phys. Lett. A, 11, 1\\
Moertsell E., Gunnarsson C., Goobar A., astro-ph/0105355 \\
Narlikar J. V., Wickramasinghe N. C., Sachs R., Hoyle F., 1997, Int. J. Mod.

\hspace{.5cm}     Phys. D, 6, 125

\noindent
Netterfield C. B. et al, astro-ph/0104460\\
Ostriker J. P., Steinhardt P. J., 1995, Nature, 377, 600\\
Overduin J. M., Cooperstock F. I., 1998, Phys. Rev. D 58, 043506\\
Ozer M., Taha M. O. 1987, Nucl. Phys. B, 287, 776\\
Perlmutter S. et al, 1999, Astrophys. J. 517, 565\\
Pryke et al, astro-ph/01044490 \\
Pskovskii Yu. P., 1977, Sov. Astron. AJ, 21, 675\\
Riess A. G. et al, 1998, Astron. J., 116, 1009\\
Riess A. G. et al, astro-ph/0104455 (to appear in Astrophys. J.)\\
Sachs R., Narlikar J. V., Hoyle F., 1996, A\&A, 313, 703\\
Sahni V., Starobinsky A., 2000, Int. J. Mod. Phys. D 9, 373\\
Schindler S., astro-ph/0107028 \\
Trimble V., 1983, Rev. Mod. Phys. 54, 1183 \\
Turner M. S., astro-ph/0106035\\
de Vaucouleurs G., Pence W. D., 1976, Astrophys. J., 209, 687\\
Vishwakarma R. G.,  2000, Class. Quantum Grav., 17, 3833\\
Vishwakarma R. G.,  2001a, Gen. Relativ. Grav., 33, No. 11 (in press) (astro-

\hspace{.5cm} ph/0106021)

\noindent
Vishwakarma R. G.,  2001b, Class. Quantum Grav. 18, 1159\\
Weller J., Albrecht A., astro-ph/0106079 \\
Wickramasinghe N. C., Wallis D. H., 1996, Astrophys. Space Sc.

\hspace{.5cm} 240, 157

\noindent
Woosley S. E., Weaver T. A. 1986, ARA \& A 24, 205 \\

\end{document}